\documentclass[aps,prl,preprint]{revtex4}
\usepackage{graphicx}

\newcommand{\ttbar}{t\bar{t}}
\newcommand{\ppbar}{p\bar{p}}

\newcommand{\met}{\mbox{$\raisebox{.3ex}{$\not$}E_T$\hspace*{0.5ex}}}
\newcommand{\scriptmet}{\mbox{\scriptsize $\raisebox{.3ex}{$\not$}E_T$\hspace*{0.5ex}}}
\newcommand{\vecmet}{\mbox{$\raisebox{.3ex}{$\not$}\vec{E}_T$\hspace*{0.5ex}}}

\begin{document}


\begin{large}
\begin{center}
Search for Anomalous Kinematics in $\ttbar$ Dilepton Events at CDF~II
\end{center}
\end{large}

\font\eightit=cmti8
\def\r#1{\ignorespaces $^{#1}$}
\hfilneg
\begin{sloppypar}
\noindent 
D.~Acosta,\r {16} J.~Adelman,\r {12} T.~Affolder,\r 9 T.~Akimoto,\r {54}
M.G.~Albrow,\r {15} D.~Ambrose,\r {43} S.~Amerio,\r {42}  
D.~Amidei,\r {33} A.~Anastassov,\r {50} K.~Anikeev,\r {15}
% ...
\end{sloppypar}
\vskip .026in
\begin{center}
(CDF Collaboration)
\end{center}

\vskip .026in
\begin{center}
\r 1  {\eightit Institute of Physics, Academia Sinica, Taipei, Taiwan 11529, 
Republic of China} \\
\r 2  {\eightit Argonne National Laboratory, Argonne, Illinois 60439} \\
\r 3  {\eightit Institut de Fisica d'Altes Energies, Universitat Autonoma
de Barcelona, E-08193, Bellaterra (Barcelona), Spain} \\
\r 4  {\eightit Istituto Nazionale di Fisica Nucleare, University of Bologna,
I-40127 Bologna, Italy} \\
\r 5  {\eightit Brandeis University, Waltham, Massachusetts 02254} \\
\r 6  {\eightit University of California at Davis, Davis, California  95616} \\
\r 7  {\eightit University of California at Los Angeles, Los 
Angeles, California  90024} \\
\r 8  {\eightit University of California at San Diego, La Jolla, California  92093} \\ 
\r 9  {\eightit University of California at Santa Barbara, Santa Barbara, California 
93106} \\ 
\r {10} {\eightit Instituto de Fisica de Cantabria, CSIC-University of Cantabria, 
39005 Santander, Spain} \\
\r {11} {\eightit Carnegie Mellon University, Pittsburgh, PA  15213} \\
\r {12} {\eightit Enrico Fermi Institute, University of Chicago, Chicago, 
Illinois 60637} \\
\r {13}  {\eightit Joint Institute for Nuclear Research, RU-141980 Dubna, Russia}
\\
\r {14} {\eightit Duke University, Durham, North Carolina  27708} \\
\r {15} {\eightit Fermi National Accelerator Laboratory, Batavia, Illinois 
60510} \\
\r {16} {\eightit University of Florida, Gainesville, Florida  32611} \\
\r {17} {\eightit Laboratori Nazionali di Frascati, Istituto Nazionale di Fisica
               Nucleare, I-00044 Frascati, Italy} \\
\r {18} {\eightit University of Geneva, CH-1211 Geneva 4, Switzerland} \\
\r {19} {\eightit Glasgow University, Glasgow G12 8QQ, United Kingdom}\\
\r {20} {\eightit Harvard University, Cambridge, Massachusetts 02138} \\
\r {21} {\eightit The Helsinki Group: Helsinki Institute of Physics; and Division of
High Energy Physics, Department of Physical Sciences, University of Helsinki, FIN-00044, Helsinki, Finland}\\
\r {22} {\eightit Hiroshima University, Higashi-Hiroshima 724, Japan} \\
\r {23} {\eightit University of Illinois, Urbana, Illinois 61801} \\
\r {24} {\eightit The Johns Hopkins University, Baltimore, Maryland 21218} \\
\r {25} {\eightit Institut f\"{u}r Experimentelle Kernphysik, 
Universit\"{a}t Karlsruhe, 76128 Karlsruhe, Germany} \\
\r {26} {\eightit High Energy Accelerator Research Organization (KEK), Tsukuba, 
Ibaraki 305, Japan} \\
\r {27} {\eightit Center for High Energy Physics: Kyungpook National
University, Taegu 702-701; Seoul National University, Seoul 151-742; and
SungKyunKwan University, Suwon 440-746; Korea} \\
\r {28} {\eightit Ernest Orlando Lawrence Berkeley National Laboratory, 
Berkeley, California 94720} \\
\r {29} {\eightit University of Liverpool, Liverpool L69 7ZE, United Kingdom} \\
\r {30} {\eightit University College London, London WC1E 6BT, United Kingdom} \\
\r {31} {\eightit Massachusetts Institute of Technology, Cambridge,
Massachusetts  02139} \\   
\r {32} {\eightit Institute of Particle Physics: McGill University,
Montr\'{e}al, Canada H3A~2T8; and University of Toronto, Toronto, Canada
M5S~1A7} \\
\r {33} {\eightit University of Michigan, Ann Arbor, Michigan 48109} \\
\r {34} {\eightit Michigan State University, East Lansing, Michigan  48824} \\
\r {35} {\eightit Institution for Theoretical and Experimental Physics, ITEP,
Moscow 117259, Russia} \\
\r {36} {\eightit University of New Mexico, Albuquerque, New Mexico 87131} \\
\r {37} {\eightit Northwestern University, Evanston, Illinois  60208} \\
\r {38} {\eightit The Ohio State University, Columbus, Ohio  43210} \\  
\r {39} {\eightit Okayama University, Okayama 700-8530, Japan}\\  
\r {40} {\eightit Osaka City University, Osaka 588, Japan} \\
\r {41} {\eightit University of Oxford, Oxford OX1 3RH, United Kingdom} \\
\r {42} {\eightit University of Padova, Istituto Nazionale di Fisica 
          Nucleare, Sezione di Padova-Trento, I-35131 Padova, Italy} \\
\r {43} {\eightit University of Pennsylvania, Philadelphia, 
        Pennsylvania 19104} \\   
\r {44} {\eightit Istituto Nazionale di Fisica Nucleare, University and Scuola
               Normale Superiore of Pisa, I-56100 Pisa, Italy} \\
\r {45} {\eightit University of Pittsburgh, Pittsburgh, Pennsylvania 15260} \\
\r {46} {\eightit Purdue University, West Lafayette, Indiana 47907} \\
\r {47} {\eightit University of Rochester, Rochester, New York 14627} \\
\r {48} {\eightit The Rockefeller University, New York, New York 10021} \\
\r {49} {\eightit Istituto Nazionale di Fisica Nucleare, Sezione di Roma 1,
University di Roma ``La Sapienza," I-00185 Roma, Italy}\\
\r {50} {\eightit Rutgers University, Piscataway, New Jersey 08855} \\
\r {51} {\eightit Texas A\&M University, College Station, Texas 77843} \\
\r {52} {\eightit Texas Tech University, Lubbock, Texas 79409} \\
\r {53} {\eightit Istituto Nazionale di Fisica Nucleare, University of Trieste/\
Udine, Italy} \\
\r {54} {\eightit University of Tsukuba, Tsukuba, Ibaraki 305, Japan} \\
\r {55} {\eightit Tufts University, Medford, Massachusetts 02155} \\
\r {56} {\eightit Waseda University, Tokyo 169, Japan} \\
\r {57} {\eightit Wayne State University, Detroit, Michigan  48201} \\
\r {58} {\eightit University of Wisconsin, Madison, Wisconsin 53706} \\
\r {59} {\eightit Yale University, New Haven, Connecticut 06520} \\
\end{center}
 
\date{\today}

\begin{abstract}
We report on a search for anomalous kinematics of 
$\ttbar$ dilepton events 
in $\ppbar$ collisions at $\sqrt{s}=1.96$~TeV
using 193~$\mbox{pb}^{-1}$ of data collected with the CDF~II detector. 
We developed a new \textit{a priori} 
technique designed to isolate the subset in a data sample 
revealing the largest deviation from standard model (SM) expectations
and to quantify the significance of this departure.
In the four-variable space considered, no particular subset shows
a significant discrepancy and we find that the probability of obtaining
a data sample less consistent with the SM than what is observed
is 1.0--4.5\%.
\end{abstract}

\pacs{14.65.Ha, 13.85.Qk, 14.80.Ly}

\maketitle


The discovery of the top quark during Run I of Fermilab's Tevatron
collider
initiated an experimental program to characterize its production and decay properties in all
possible decay channels.  Within the standard model (SM) the top quark decays almost exclusively
to a $W$ boson and a bottom quark; the ``dilepton'' decay channel here denotes the
case where the two $W$ bosons
from a $\ttbar$ pair
both decay into final states containing an
electron or a muon, accounting for about 7\% of all SM $\ttbar$ decays.
These events are characterized by two energetic leptons, two jets from
the hadronization of the bottom quarks, and large missing
energy from the unobserved neutrinos.
The CDF and D\O\ Collaborations' measurements of the $\ttbar$ production
cross section in the dilepton channel in Run~I~\cite{Abe:1998iz}
showed a slight excess over SM predictions~\cite{Bonciani:1998vc}.
Perhaps more interestingly, several of the events observed in
the Run I data had
missing transverse energy (\met)
and lepton $p_{T}$'s~\cite{coord} large enough to call into
question their compatibility with SM top decay kinematics.  In fact, it
was suggested that the kinematics of these events could be better described
by the cascade decays of heavy squarks~\cite{Barnett:1996hw}, compelling us
to subject the top dilepton sample to careful scrutiny in Run~II.

In a previous Letter~\cite{Acosta:2004uw}, we reported a measurement of
the $\ttbar$ production cross section
in the dilepton channel at Run~II and found good agreement with the SM
expectation.
Here we present the results of a detailed
analysis of the kinematics of that data sample.
Motivated by the possible anomalies in the top Run~I dilepton
sample, we devised a search for new physics based on the
comparison of kinematic features of observed events with
those expected from the SM, assuming a 175~GeV/$c^{2}$ top mass~\cite{Demortier:1999vv}.
The search is designed to be sensitive
to any physical process that gives rise to events with specific kinematics
different from those expected from SM top and backgrounds, especially
processes that result in kinematics similar to the aforementioned
Run~I events.  The method seeks to isolate the subset of events in a data sample
with the largest concentration of possible non-SM physics
and to assign a
probability that quantifies its departure from the SM.

Reference~\cite{Acosta:2004uw} provides a description of the CDF-II detector,
the event selection, and the data and simulation samples used for this analysis~\cite{dil}.
The basic selection requirements are
(i) two oppositely-charged, well-identified leptons
($e$ or $\mu$) with $p_{T}~>~20$~GeV/$c$, (ii) at least two jets
with $E_{T}~>~15$~GeV, and (iii) \met~$>$~25~GeV.
Several other topological requirements are
made to further purify the sample and are detailed in~\cite{Acosta:2004uw}.
With this selection, the SM predicts a yield of $8.2 \pm 1.1$ $\ttbar$ events
(assuming a $\ttbar$ cross section of 6.7~pb~\cite{Bonciani:1998vc}),
and $2.7 \pm 0.7$ events from other SM processes
(mainly production of dibosons, $W$~+~associated jets, and Drell-Yan
events) in our sample.  Thirteen events are observed.

We consider
a minimal set of assumptions about the nature of possible non-SM physics
in order to make an \textit{a priori} choice of which kinematic quantities
to investigate.
The Tevatron provides us with the opportunity to look for
phenomena beyond the presently known mass spectrum.
This together with the hints from the Run~I data sample leads us to focus our
search on events with large lepton $p_{T}$ and large \met resulting from
the decay of an unknown heavy particle.
In addition, two-body decays of massive particles (\textit{e.g.} heavy
chargino decay $\tilde{\chi}^\pm \to \ell^{\pm}\tilde{\nu}$) tend to result in
topologies where the charged lepton and the \met direction are
back-to-back, whereas this tends not to be the case for the SM $\ttbar$
dilepton signature.
Thus we expect the following variables to be sensitive to a wide range of new physics:
the event's \met, 
the transverse momentum of the leading (\textit{i.e.} highest-$p_{T}$) lepton $p_T^\ell$,
and the 
angle $\Phi_{\ell m}$
between the leading lepton and the direction of the \met in 
the plane transverse to the beam.

We define an additional kinematic variable as follows.
The initial and intermediate state particles in the $\ttbar$ decay
impose constraints on the
final state product properties, $m(\ell_1\nu_1) = m(\ell_2\nu_2) = m_W$ and  
$m(\ell_1\nu_1 b_1) = m(\ell_2\nu_2 b_2) = m_t = 175$~GeV/$c^{2}$.  These four
constraints leave two of the six unknown neutrino momentum components
unspecified when solving the system of kinematic equations.  To fully
reconstruct the event, we scan over these two remaining degrees of freedom and
compare the resulting neutrino momentum sum ($\vecmet^{pred}$) with the
\vecmet measured in the event ($\vecmet^{obs}$) by computing
\begin{equation}
{\mathcal T}(\vecmet^{pred}) = \exp\left\{ -\left|{\vecmet}^{pred} -
{\vecmet^{obs}}\right|^2/2\sigma_{\scriptmet}^2\right\}
\end{equation}
where $\sigma_{\scriptmet}$ parameterizes uncertainty on $\met$ due to
mismeasurement of the underlying event.
When performing the scan
we assume detector resolutions to be Gaussian for the lepton and jet momenta and
smear the observed values accordingly;
the $\vecmet^{pred}$ value is then recomputed according to the smeared jet and
lepton energies.  We define a variable $T$ as the square root of the
integral of ${\mathcal T}$ over the possible values of $\vecmet^{pred}$ determined
from the scan and summed over a two-fold ambiguity in the lepton-b-jet pairing.
This
variable $T$ represents how well an event's kinematics satisfy the $\ttbar$ dilepton
decay hypothesis;
a non-$\ttbar$ dilepton event has on average a small value of $T$ compared to
$\ttbar$ events.

As mentioned before, we concentrate our search on events with large values
of \met, $p_{T}^{\ell}$, and $\Phi_{\ell m}$ and small values of $T$.  We
therefore assign the following weight to each event:
\begin{equation}
W = (w_{\scriptmet} \cdot w_{p_{T}^{\ell}} \cdot w_{\Phi_{\ell m}} \cdot w_T )^{1/4}
\end{equation}
where $w_{\scriptmet}, w_{p_{T}^{\ell}}, w_{\Phi_{\ell m}},$ and $w_T$ represent
probabilities (assuming the SM) for an
event to have a $\met, p_{T}^{\ell}, \Phi_{\ell m}$ larger than that observed and 
a $T$ smaller than that observed, respectively.
We then construct 13 subsets (``$K$-subsets'') of the data;
the first subset ($K=1$) contains
only the event with the lowest weight $W$, the second subset ($K=2$)
contains only the two events with the two lowest weights, and so on.

To quantify the departure of the $K$-subsets from the SM
predictions we do 
a shape comparison using the Kolmogorov-Smirnov (KS) 
statistic~\cite{KS_test}.
For each of the four variables $i$, the KS deviation
$\Delta_{K,i}$ between the SM 
cumulative function and the cumulative function of the
$K$-subset is computed. 
To assess the probability of this deviation we generate
100,000 pseudoexperiments by randomly drawing events from large Monte
Carlo samples of $\ttbar$ and SM backgrounds. 
The number of events corresponding to each SM process is sampled from a Poisson
distribution with mean equal to the number of events expected after event
selection. Only
pseudoexperiments with a total of 13 events are accepted. Further, in each 
pseudoexperiment, $K$-subsets are formed and the respective $\Delta_{K,i}$ 
for each are calculated.
We thus build probability distribution functions for
$\Delta_{K,i}$ from which the KS probability $p_{K,i}$ can be computed.
Next we calculate the geometric mean $\Pi_{K}$ of the four $p_{K,i}$'s for each
pseudoexperiment and form
the probability distribution functions ${\cal F}_K$ 
such that
the quantity
\begin{equation}
    P_{K} = \int_{0}^{\Pi_{K}^{obs}}  {\cal F}_K(\Pi)\,d\Pi
\end{equation}
determines how well each $K$-subset agrees with the SM
expectation based on the combined information from the four variables.
We define $Q$ as the value of $K$ with the smallest $P_K$.
By isolating this ``unlikely'' subset $Q$ (where ``unlikely'' here denotes
having large $p_{T}^{\ell}$, \met, $\Phi_{\ell m}$ and/or
small $T$),
we minimize
the dilution of a possible signal from the inclusion of SM events.

We use the quantity $P_Q$ as the test statistic to quantify the 
discrepancy of the data with the SM.
Generating another set of 100,000 pseudoexperiments from SM Monte Carlo
and repeating the above procedure, we determine $P_Q$ for each pseudoexperiment
and build the probability distribution function ${\cal L}(P_Q)$ such that
the significance of departure of the $Q$-subset of events from the SM is
\begin{equation}
  \alpha = \int_0^{ P_{Q}^{data}} {\cal L}(P_Q)\,dP_Q.
\end{equation}
$\alpha$ is the $p$-value of the test, representing the probability to
obtain a data sample less consistent with the SM than what is actually observed.
Sufficiently low values of $\alpha$ would indicate the presence of new
physics in the data sample,  and the $Q$ events would represent the subsample
of the data with the largest concentration of new physics.

In order to evaluate the performance of the method, we simulated a sample of
squark decays using \textsc{PYTHIA}~\cite{Sjostrand:2000wi} and the SUSY parameters
suggested in~\cite{Barnett:1996hw}.
As a performance benchmark, we construct a 50\%:50\% mixture of the SM and SUSY
and ask how often we would observe a $p$-value ($\alpha$) less than 0.3\% (the
equivalent of a $3\sigma$ effect) when 13-event
pseudoexperiments are drawn from this sample.  We find that $\approx50\%$ of these
pseudoexperiments yield $\alpha < 0.3\%$.  Moreover, the concentration of SUSY
events in the most unlikely $K$-subset found is on average 80\%.  By contrast,
a KS test without using subsamples finds $\alpha < 0.3\%$ only 21\% of the time
and does not isolate 
a mostly-SUSY subset. 

We test our procedure as well as our ability to 
correctly 
simulate our kinematic variables
in a high-statistics control sample of 973 $W+\geq3$~jets events. 
We compare these data with a Monte Carlo simulation of \met, $p_{T}^{\ell}$ and
$\Phi_{\ell m}$ using 
$W+$~associated jet, QCD, and $\ttbar$ production processes
added in the amounts expected from the SM.
We apply a 3-dimensional
version of our technique and observe that the data have a
high $p$-value ($\alpha = 35.1\%$), indicating good modeling of the
data by the simulation.
 
We test the modeling of $T$ in a control sample of $W+4$~jets
events,
treating the leading jet as a second lepton and the subleading jet as
a second neutrino.
We apply this reconstruction to the data and to an appropriately weighted
sample of simulated $\ttbar$ and \textsc{ALPGEN+HERWIG} $W+4$~parton Monte Carlo~\cite{Mangano:2002ea}.
We observe a KS probability
of 0.97 for the respective $T$ distributions, indicating good agreement between
simulation and the data.

\begin{figure*}
\includegraphics[width=0.45\textwidth]{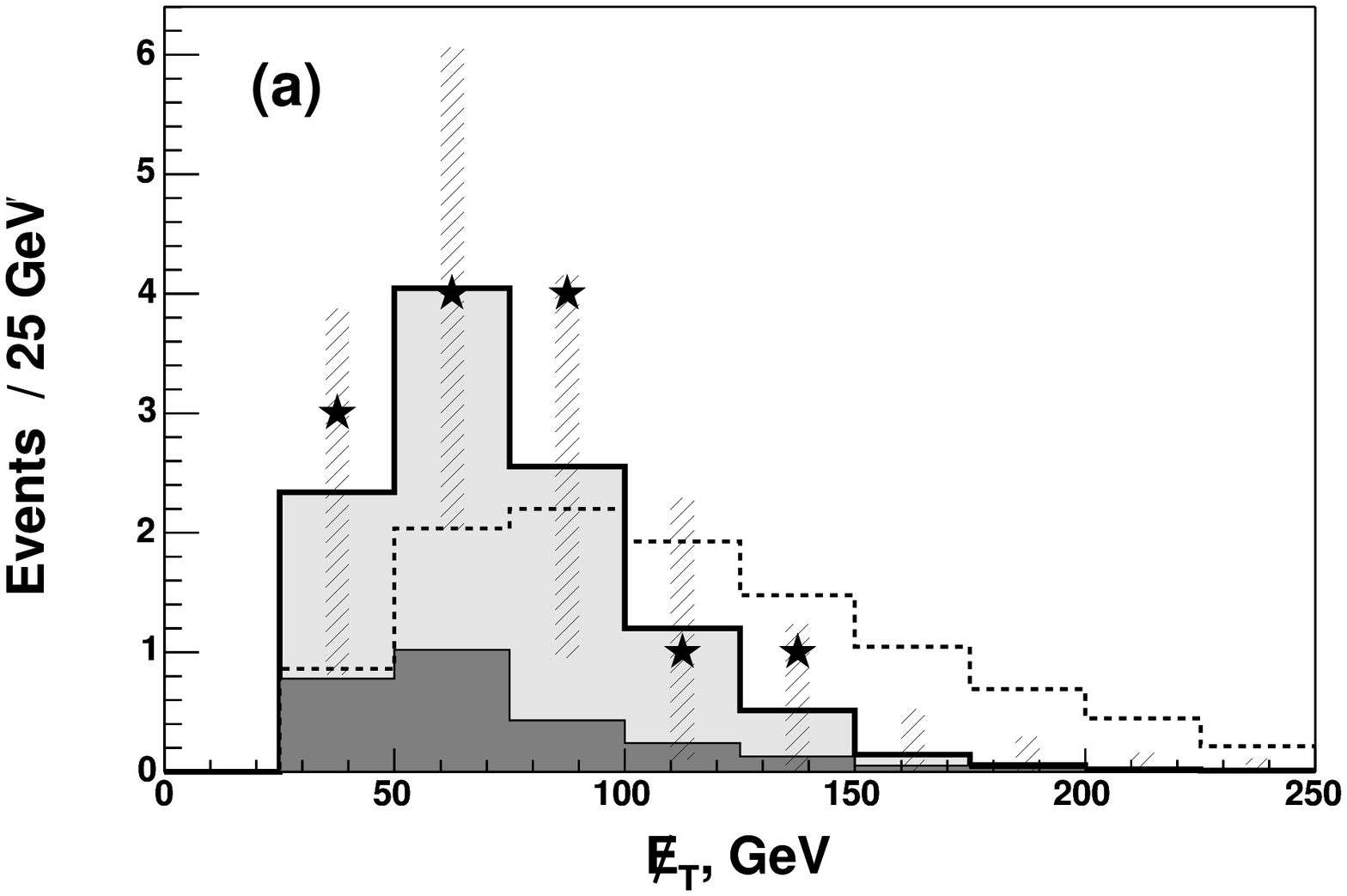}
\includegraphics[width=0.45\textwidth]{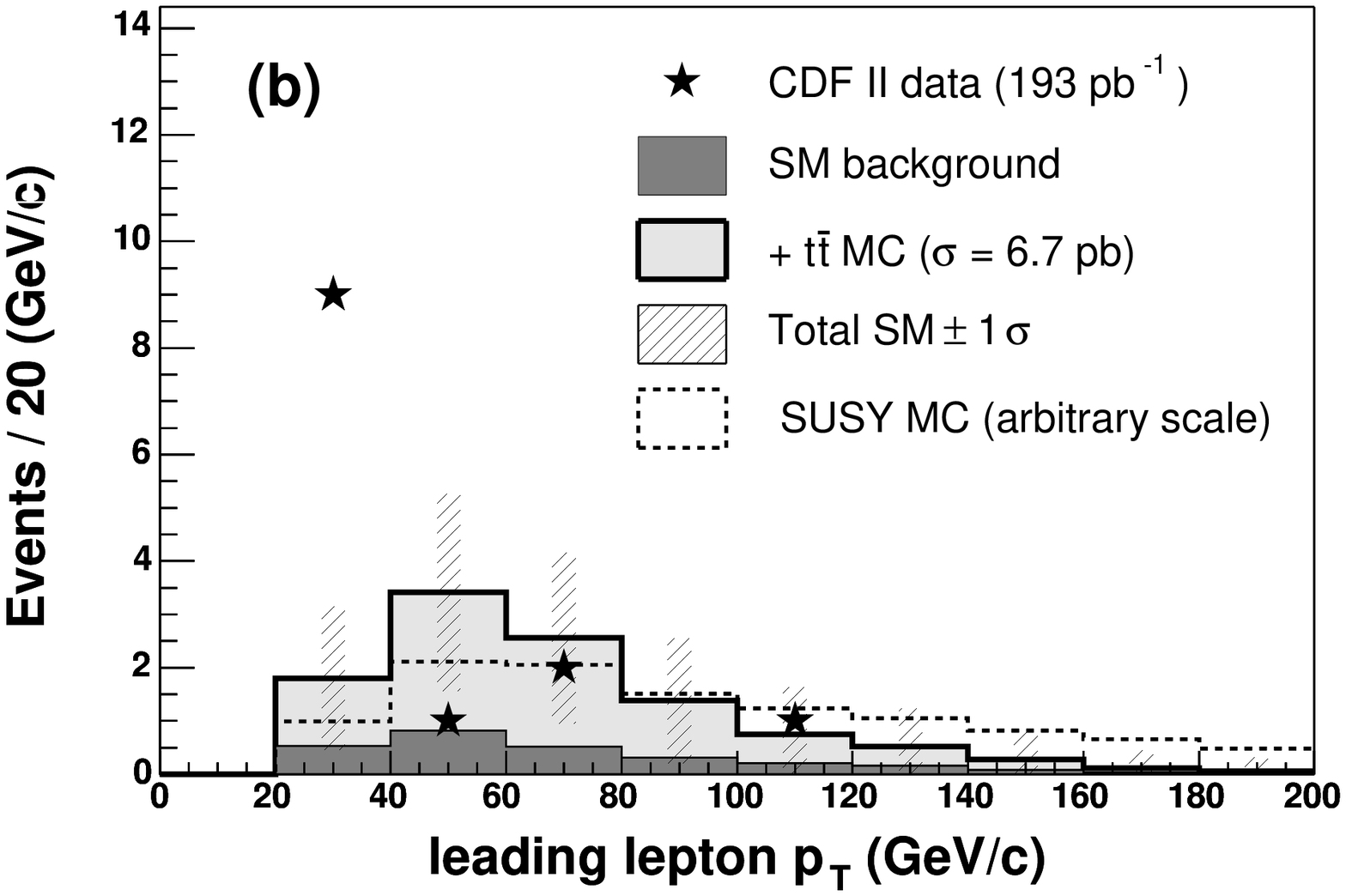}\\
\includegraphics[width=0.45\textwidth]{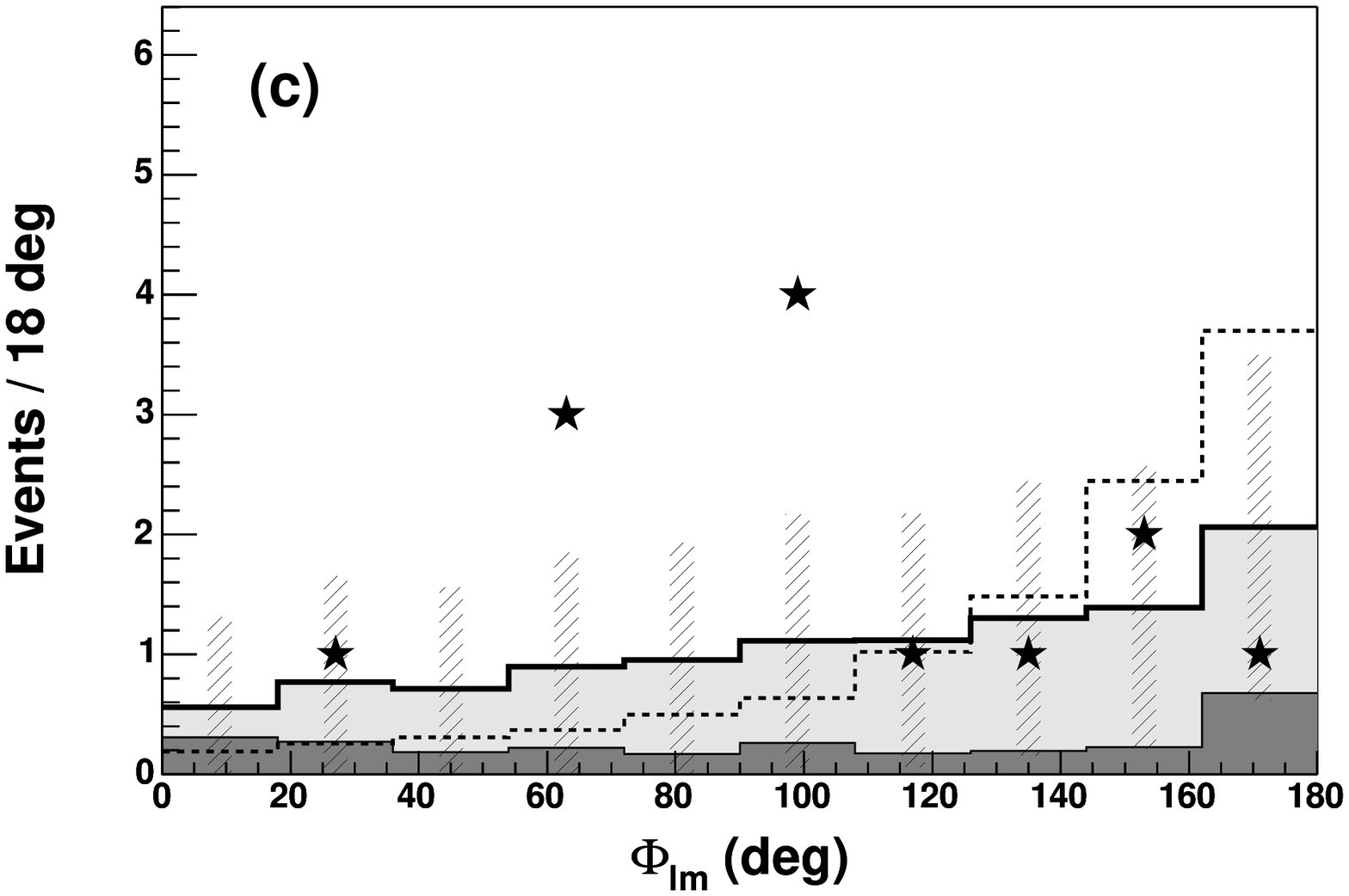}
\includegraphics[width=0.45\textwidth]{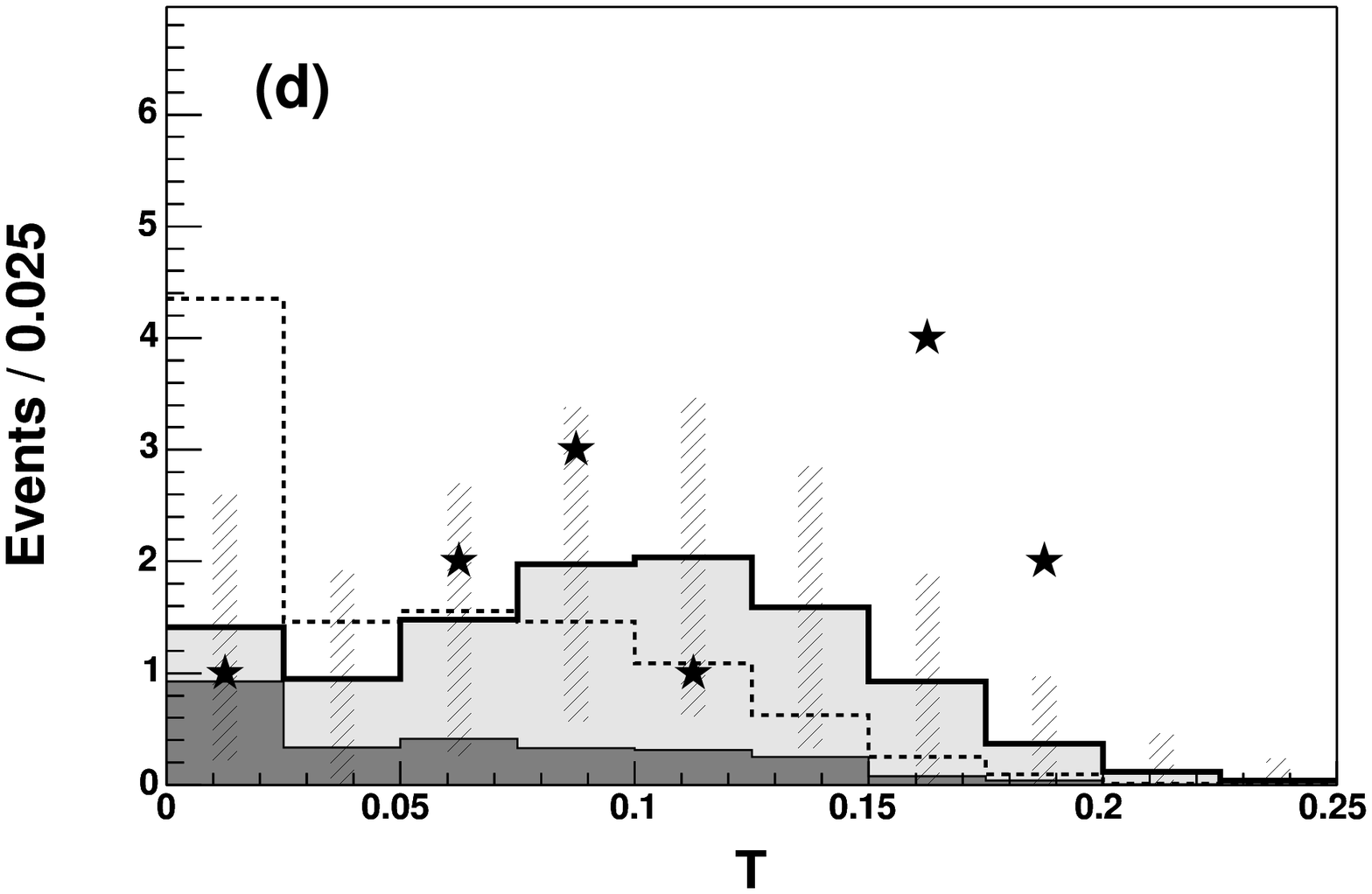}
\caption{\label{fig:dil1}
\protect\met, leading lepton $p_T$, $\Phi_{\ell m}$,
and $T$ distributions for the top dilepton sample.  The hatched regions represent the
Poisson uncertainty on the expectation in a given bin.  The dashed histograms are
the expected distributions from the SUSY MC described in the text.
}
\end{figure*}

Having established that data are adequately modeled by the simulation, we
apply the outlined technique to the $\ttbar$ dilepton sample.
The distributions of the
selected variables for $\ttbar$ dilepton events are presented 
in Figure~\ref{fig:dil1}.
We find the most unlikely subset of events to
be the entire data set (\textit{i.e.} $Q=13$), with a
$p$-value = 1.6\%.
This result is entirely driven by the excess of leptons at low $p_{T}$ ($<40$~GeV/$c$) 
seen in Figure~\ref{fig:dil1}b; since
the method effectively orders the subsets from high $p_{T}$ to low $p_{T}$, the
$p$-value decreases as more of the low-$p_{T}$ excess is included, reaching a
minimum when the entire data sample is considered.

\begin{figure}
\includegraphics[width=0.45\textwidth]{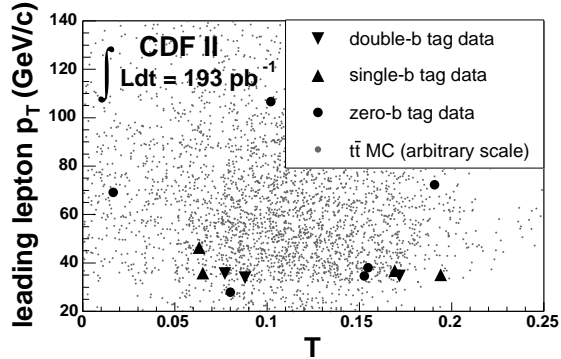}
\caption{\label{fig:btags}
Top dilepton events in ($p_{T}^{\ell}$, $T$) plane with b-tagging information.
}
\end{figure}

A natural question to ask about the low-$p_{T}$ events
is whether they can be attributed to underestimated non-$\ttbar$ SM backgrounds.  To address this,
we used a
displaced secondary vertex ``$b$-tag'' algorithm~\cite{Acosta:2004hw} to look for long-lived
$b$-hadron decays in the events; the fraction of non-$\ttbar$ SM dilepton
events containing bottom quarks is expected to be negligible. 
We present the $b$-tag content of the
sample as well as the distribution of events in the ($p_{T}^{\ell}$, $T$) plane in
Figure~\ref{fig:btags}.  We note that six of the nine low-$p_{T}$
events contain at least
one identified $b$-jet.  We also note that more than half of the low-$p_{T}$ events
are consistent with the $\ttbar$
kinematic hypothesis with large values of $T$, as opposed
to the small values of $T$ ($<0.05$) favored by non-$\ttbar$ SM backgrounds (see
Figure~\ref{fig:dil1}d).
We thus conclude
that the low-$p_{T}$ events are not likely to have arisen from
non-$\ttbar$ SM processes; details of the thirteen events can be found
elsewhere~\cite{andrew_thesis}.

We next evaluate the effect of systematic uncertainties.
Uncertainties in the shapes of kinematic distributions from
sources listed in Table~\ref{tab:syu} lead to
an uncertainty in the
probability distribution function 
${\cal L}(P_Q)$, and consequently to an uncertainty in the
significance level of our measurement.
We consider each source of systematic uncertainty and build a new
probability distribution function ${\cal L}^\prime(P_Q)$.
We then determine a new
$p$-value $\alpha^\prime$ via
\begin{equation}
  \alpha^\prime = \int_0^{ P_{Q}^{data}} {\cal L}^\prime(P_Q)\,dP_Q.
\end{equation}
Table~\ref{tab:syu} shows the values of $\alpha^{\prime}$ obtained for
different sources of uncertainty.  Generating an ${\cal L}^\prime(P_Q)$
with the inclusion of all systematic effects that give a $p$-value greater than that
observed in the data (1.6\%)
results in a maximum $p$-value of 4.5\%; a minimum $p$-value of 1.0\% is obtained when
a background estimate $1\sigma$ lower than nominal is used.  All other combinations
of systematic effects result in $p$-values lying within this range.

In conclusion, we have assessed the consistency of the $\ttbar$ dilepton sample
with the SM in the four-variable space described and find a $p$-value of
1.0--4.5\%.  Our method is designed to be especially sensitive to
data subsets that preferentially populate regions where new high-$p_{T}$ physics can be
expected.  No such subset was found in our data. 
We have noted that the lepton $p_{T}$ distribution exhibits a mild excess at low $p_{T}$;
however, it can be concluded that new physics
scenarios invoked to describe the high-$p_{T}^{\ell}$/high-\met events observed
in Run~I are not favored by the current Run~II data.

\begin{table}
\caption{\label{tab:syu}
$p$-values obtained upon inclusion of systematic effects.  The last row shows
the maximum range of $p$-values resulting from various combinations of the individual
systematics.}
\begin{ruledtabular}
\begin{tabular}{cc}
Source of uncertainty                    &  $\alpha^\prime$ (\%)   \\ \hline
MC generator                             &      1.6 \\  
Initial (final) state radiation          &      1.2 (1.6)\\
Parton distribution functions            &      1.9 \\
$M_{top}$ = 170 (180) GeV                &      1.4 (2.1)\\
Jet energy scale, +1 (-1) $\sigma$       &      2.1 (2.6)\\              
Background estimates, +1 (-1) $\sigma$   &      2.7 (1.0)\\
\hline
combined                                &  1.0--4.5\\
\end{tabular}
\end{ruledtabular}
\end{table}

We thank the Fermilab staff and the technical staffs of the participating institutions for their vital contributions. 
This work was supported by the U.S. Department of Energy and National Science Foundation; the Italian Istituto Nazionale 
di Fisica Nucleare; the Ministry of Education, Culture, Sports, Science and Technology of Japan; the Natural Sciences and 
Engineering Research Council of Canada; the National Science Council of the Republic of China; the Swiss National Science 
Foundation; the A.P. Sloan Foundation; the Bundesministerium fuer Bildung und Forschung, Germany; the Korean Science and 
Engineering Foundation and the Korean Research Foundation; the Particle Physics and Astronomy Research Council and the 
Royal Society, UK; the Russian Foundation for Basic Research; the Comision Interministerial de Ciencia y Tecnologia, 
Spain; the Research Corporation; and in part by the European Community's Human Potential Programme under contract
HPRN-CT-2002-00292, Probe for New Physics.

\bibliography{pks_prl}

\begin{thebibliography}{17}
\bibitem[{coo()}]{coord}
\bibinfo{note}{We use a cylindrical coordinate system about the beam axis in
  which $\theta$ is the polar angle. We define transverse momentum
  $p_{T}=p\sin{\theta}$ and transverse energy $E_{T}$ similarly. Missing
  transverse energy (\vecmet) is defined as $-\sum_{i}E_{T}^{i}\hat{n}_{i}$,
  where $\hat{n}_{i}$ is the unit vector in the azimuthal plane that points
  from the $p\bar{p}$ interaction region to the $i^{th}$ calorimeter tower.
  \vecmet\ is further corrected for the energy and momentum of identified
  muons.}
\bibitem[{dil()}]{dil}
\bibinfo{note}{Two selections are described in the cited reference. In order to
  minimize sensitivity to non-$\ttbar$ SM processes, we use the higher-purity
  selection referred to as ``DIL'' therein.}
\end{thebibliography}

\end{document}